\documentclass[aps,prb,twocolumn,superscriptaddress,showpacs]{revtex4-1}

\usepackage{graphicx}
\usepackage{calc}
\usepackage{bm}
\usepackage{color}

\begin{document}

\title{Signature of Fermi arc surface states in Andreev reflection at the WTe$_2$ Weyl semimetal surface.}

\author{A.~Kononov}
\affiliation{Institute of Solid State Physics of the Russian Academy of Sciences, Chernogolovka, Moscow District, 2 Academician Ossipyan str., 142432 Russia}
\author{O.O.~Shvetsov}
\affiliation{Institute of Solid State Physics of the Russian Academy of Sciences, Chernogolovka, Moscow District, 2 Academician Ossipyan str., 142432 Russia}
\affiliation{Moscow Institute of Physics and Technology, Institutsky per. 9, Dolgoprudny, 141700 Russia}
\author{S.V.~Egorov}
\affiliation{Institute of Solid State Physics of the Russian Academy of Sciences, Chernogolovka, Moscow District, 2 Academician Ossipyan str., 142432 Russia}
\author{A.V.~Timonina}
\affiliation{Institute of Solid State Physics of the Russian Academy of Sciences, Chernogolovka, Moscow District, 2 Academician Ossipyan str., 142432 Russia}
\author{N.N.~Kolesnikov}
\affiliation{Institute of Solid State Physics of the Russian Academy of Sciences, Chernogolovka, Moscow District, 2 Academician Ossipyan str., 142432 Russia}
\author{E.V.~Deviatov}
\affiliation{Institute of Solid State Physics of the Russian Academy of Sciences, Chernogolovka, Moscow District, 2 Academician Ossipyan str., 142432 Russia}

\date{\today}

\begin{abstract}
We  experimentally investigate charge transport through the interface between a niobium superconductor and a three-dimensional WTe$_2$ Weyl semimetal. In addition to classical Andreev reflection, we observe sharp non-periodic subgap resistance resonances. From an analysis of their positions, magnetic field and temperature dependencies, we can interpret them as an analog of Tomasch oscillations for transport along the topological surface state across the region of proximity-induced superconductivity at the Nb-WTe$_2$ interface.  Observation of distinct geometrical resonances implies a specific transmission direction for carriers, which is a hallmark of the Fermi arc surface states.
\end{abstract}


\maketitle

\section{Introduction}

Recent interest to Weyl semimetals is mostly connected with topological surface properties~\cite{armitage}.  Weyl semimetals are conductors which, like other topological materials~\cite{hasan,zhang,das,chiu}, are characterized by topologically protected conducting surface states.  
The concept of Weyl semimetals has been extended to type II materials~\cite{armitage}, like MoTe$_2$ and WTe$_2$, where constant energy surfaces are open electron and hole pockets with a Weyl point at their touching. Weyl points are  topologically protected and their projections on the surface Brillouin zone are connected by Fermi arc surface states. For these materials, surface states were demonstrated in several experiments~\cite{wang,wu}, although their topological nature is stil debatable~\cite{bruno,li}. In contrast to three-dimensional  topological insulators~\cite{hasan} described by $Z_2$ invariant, Weyl surface states  inherit the chiral property of the Chern insulator edge states~\cite{armitage}.  

Topological materials  exhibit non-trivial physics in  proximity with a superconductor~\cite{SNS1,SNS2,kundu}. For a single normal-superconductor (NS) contact, Andreev reflection~\cite{andreev} allows low-energy electron transport from normal metal to superconductor  by creating  a Cooper pair, so a hole is reflected back to the normal side of the junction~\cite{tinkham}. The process can be more complicated~\cite{heslinga}  if Andreev transport goes through an intermediate conductive region, e.g., the topological surface state at the NS interface~\cite{klapwijk17,nbsemi,ingasb}. Also, geometrical resonances are predicted~\cite{adroguer,melnikov} within the topological surface state in proximity with a superconductor, analogous to classical Tomasch~\cite{tomasch1,tomasch2,tomasch_exp1,tomasch_exp2} effect. 

allows only internodal transport at the junction of the Weyl semimetal with a superconductor

Here, we experimentally investigate charge transport through the interface between a niobium superconductor and a three-dimensional WTe$_2$ Weyl semimetal. In addition to classical Andreev reflection, we observe sharp non-periodic subgap resistance resonances. From an analysis of their positions, magnetic field and temperature dependencies, we can interpret them as an analog of Tomasch oscillations for transport along the topological surface state across the region of proximity-induced superconductivity at the Nb-WTe$_2$ interface.  Observation of distinct geometrical resonances implies a specific transmission direction for carriers, which is a hallmark of the Fermi arc surface states.

\section{Samples and technique}

\begin{figure}
\includegraphics[width=0.7\columnwidth]{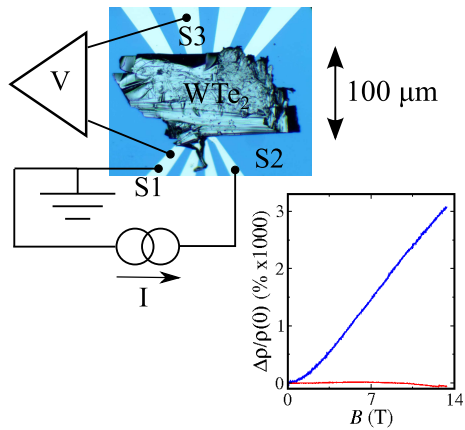}
\caption{(Color online) Sketch of the sample with niobium contacts to the bottom surface of a WTe$_2$ crystal.  70~nm thick  niobium superconducting leads  are formed on the insulating SiO$_2$ substrate. A WTe$_2$ single crystal is weakly pressed to the niobium leads pattern, forming planar Nb-WTe$_2$ junctions with $\approx  10\times 10 \mu\mbox{m}^2$ area.  Charge transport is investigated in a standard three-point technique: the studied contact (S1) is grounded and two other contacts (S2 and S3) are used for applying current (below 100~$\mu$A) and measuring WTe$_2$ potential.  Inset demonstrates large positive magnetoresistance $\rho(B)-\rho(B=0)/\rho(B=0)$   for our WTe$_2$ samples  at 1.2~K in normal magnetic field (the blue curve), which goes to zero in parallel one (the red curve), as it has been shown for WTe$_2$ Weyl semimetal~\protect\cite{mazhar}. The current is parallel to the $a$ axis of WTe$_2$.
}
\label{sample}
\end{figure}

WTe$_2$ compound was synthesized from elements by reaction of metal with tellurium vapor in the sealed silica ampule. The WTe$_2$ crystals were grown by the two-stage iodine transport~\cite{growth1}, that previously was successfully applied~\cite{growth1,growth2} for growth of other metal chalcogenides like NbS$_2$ and CrNb$_3$S$_6$. The WTe$_2$ composition is verified by energy-dispersive X-ray spectroscopy. The X-ray diffraction (Oxford diffraction Gemini-A, MoK$\alpha$) confirms $Pmn2_1$ orthorhombic single crystal WTe$_2$ with lattice parameters $a=3.48750(10)$~\AA, $b= 6.2672(2)$~\AA, and $c=14.0629(6)$~\AA.

A sample sketch is presented in Fig.~\ref{sample}. We use dc magnetron sputtering to deposit a 70~nm thick niobium film on the insulating SiO$_2$ substrate. Superconducting leads are formed by lift-off technique. A WTe$_2$ single crystal ($\approx 0.5 \mbox{mm}\times 100\mu\mbox{m} \times 0.5 \mu\mbox{m} $ dimensions) is  weakly pressed to the niobium leads pattern, so the planar Nb-WTe$_2$ junctions (with $\approx  10\times 10 \mu\mbox{m}^2$ area) are formed at the bottom surface of the WTe$_2$ crystal. 

\begin{figure}
\includegraphics[width=\columnwidth]{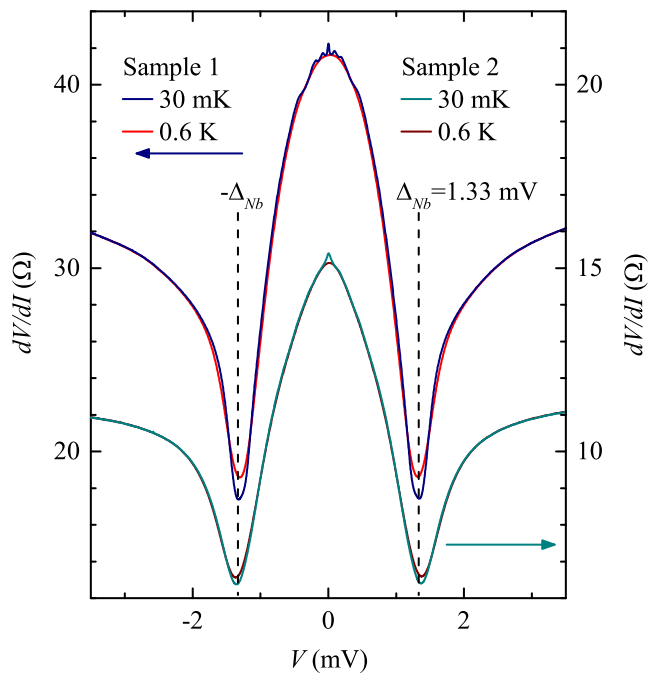}
\caption{(Color online)  Examples of $dV/dI(V)$ characteristics for two Nb-WTe$_2$ junctions for two different temperatures. The niobium superconducting gap $\Delta_{Nb}\simeq\pm$~1.33~mV is marked by dashed lines. The general behavior of $dV/dI(V)$ curves is not sensitive to temperature for $T<<T_c=9$~K. }
\label{IVa}
\end{figure}

We study electron transport across a single Nb-WTe$_2$ junction in a standard three-point technique, see Fig.~\ref{sample} (c): the studied contact is grounded and two other contacts are used for applying current (below 100~$\mu$A) and measuring WTe$_2$ potential. To obtain $dV/dI(V)$ characteristics,   the dc current is additionally modulated by a low ac (0.5~$\mu$A, 1.2~kHz) component. We measure both,  dc ($V$) and ac ($\sim dV/dI$) components of the WTe$_2$ potential by using a dc voltmeter and a lock-in, respectively. We check, that the lock-in signal is independent of the modulation frequency in the 500 -- 2500~Hz range, which is defined by applied ac filters.

 	  To extract features specific to  WTe$_2$ Weyl semimetal surface states, the measurements are performed in a dilution refrigerator covering 30~mK--1.2~K temperature range. We check by standard magnetoresistance measurements that our WTe$_2$ samples demonstrate large, non-saturating positive magnetoresistance $\rho(b)-\rho(B=0)/\rho(B=0)$  in normal magnetic field, which goes to zero in parallel one, see inset to Fig.~\ref{sample},  as it has been shown for WTe$_2$ Weyl semimetal~\protect\cite{mazhar}.

\section{Experimental results}

\begin{figure}
\includegraphics[width=\columnwidth]{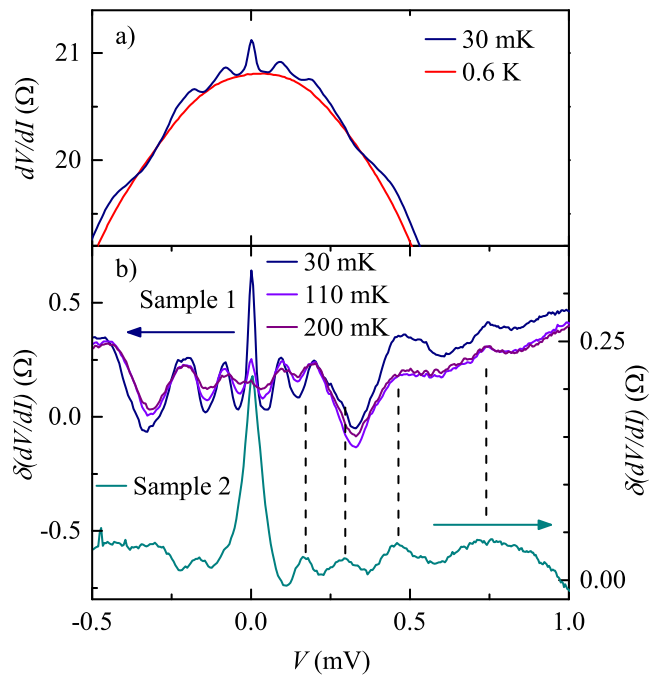}
\caption{(Color online) (a)  The central part of $dV/dI(V)$ curves at 30~mK and 0.6~K temperatures. Specifics of Nb-WTe$_2$ junction appears in subgap resistance resonances below 0.6~K.
  (b) The difference $\delta (dV/dI(V))$ between $dV/dI(V)$ characteristics at low temperatures (30~mK, 110~mK, 200~mK) and after vanishing of resonances at 0.6~K. The positions of the resonances are denoted by dashed lines, they are  non-periodic and concentrated strictly within the superconducting gap. $dV/dI$ is a maximum at zero bias for both Nb-WTe$_2$ junctions.}
\label{IVb}
\end{figure}

Examples of $dV/dI(V)$ characteristics are shown in Fig.~\ref{IVa}  for two different junctions with  different $R_N$. The obtained $dV/dI(V)$ curves are verified to be independent of the mutual positions of the current/voltage, so they only reflect the transport parameters of the grounded Nb-WTe$_2$  junction. 

 The main $dV/dI(V)$ behavior is consistent with the standard one~\cite{tinkham} of a single  Andreev  junction: every curve demonstrates a clearly defined  superconducting gap $\Delta_{Nb}\simeq\pm$~1.33~mV (denoted by dashed lines), which  is in a good correspondence with the expected $T_c\approx 9$~K for niobium. The the subgap resistance   is undoubtedly finite, which is only possible due to Andreev reflection. 
	It exceeds the normal resistance value, so  single-particle scattering is significant at the Nb-WTe$_2$ interface~\cite{tinkham}. The interface scattering is expected, since the sputtered niobium is natively oxidized prior to placing a WTe$_2$ single crystal onto the Nb leads pattern. A transmission of the interface $T$ can be estimated as $\approx 0.73-0.76$ for these junctions, which corresponds to the BTK barrier strength~\cite{tinkham} $Z\approx 0.6$.

Specifics of the WTe$_2$ Weyl semimetal appears as sharp subgap resistance resonances, see Figs.~\ref{IVa}  and ~\ref{IVb}, which can not be expected~\cite{tinkham} for  single Andreev NS contact. The resonances are suppressed completely above 0.6~K. Since the general behavior of $dV/dI(V)$ curves is not sensitive to temperature much below $T_c=9$~K, see Fig.~\ref{IVa}, the resonances can be  analyzed in detail by subtracting the high-temperature (0.6~K) monotonous $dV/dI$ curve from the low-temperature (30~mK) one. The result ($\delta (dV/dI(V))$) is shown in Fig.~\ref{IVb} (b) for the junctions from Fig.~\ref{IVa}, the positions of the resonances are denoted by dashed lines. They are concentrated strictly within the superconducting gap and  $dV/dI$ is a maximum at zero bias, the distance between the resonances is obviously  increasing with bias in Fig.~\ref{IVb} (b).  The number of visible resonances is different for different junctions.  With increasing temperature, the amplitude of every resonance is diminishing, while its position is invariant, see Fig.~\ref{IVb} (b).

  $dV/dI(V)$ curves are shown in Fig.~\ref{IV_magn} (a) for different magnetic fields, oriented along a-axis of the $WTe_2$ crystal. In contrast to temperature suppression, the zero-bias $dV/dI$ peak survives above  0.85~T, while the resonances disappear completely. Moreover,  $dV/dI(V)$ curve still demonstrates non-trivial shape even for higher fields, with non-zero first derivative at zero bias, which also differs from temperature  $dV/dI(V)$ suppression.  
This  evolution is shown in Fig.~\ref{IV_magn} (b) as $\delta (dV/dI(V))$ colormap. The oscillations' positions are constant up to their disappearance at  0.85~T, while superconductivity is completely suppressed around 4~T magnetic field in our samples.

\section{Discussion} \label{disc}

As a result, we observe pronounced subgap $|eV|<\Delta_{Nb}$ resonances for well-developer Andreev  $dV/dI(V)$ curves for a single Nb-$WTe_2$ SN junction. In general, resonance conditions require particle propagation between two different (SN or NN) interfaces. 

For a  thin $WTe_2$ crystal, it is naturally to think about vertical transport, normal to the Nb-$WTe_2$ interface.  In this case, both  Tomasch~\cite{tomasch1,tomasch2} and MacMillan-Rowell~\cite{mcmillan1,mcmillan2} geometrical resonances could be anticipated, which originates~\cite{tomasch1,mcmillan1,mcmillan2,tomasch_exp1,tomasch_exp2} due to the space restriction in the S or N regions, respectively.  However, these geometrical resonances should be observed at energies above the niobium superconducting gap, which contradicts to the experimental observations in  Fig.~\ref{IVa}. Also, MacMillan-Rowell oscillations~\cite{mcmillan1,mcmillan2,osbite} for bulk carriers in the WTe$_2$ crystal are strictly periodic, which is is obviously not the case in  Fig.~\ref{IVb}. 

Another possibility is multiple Andreev reflections~\cite{tinkham} (MAR), but the experimental $dV/dI(V)$ curves are invariant to the choice of the potential contact. The resonance positions in Fig.~\ref{IVb} (b) does not correspond to the MAR sequence $E_n=2\Delta_{Nb}/n, n=1,2,3...$. Significant single-particle scattering at the Nb-WTe$_2$ interface is  inconsistent with MAR observation conditions for highly separated (more then 5~$\mu$m) Nb contacts.  

\begin{figure}
\includegraphics[width=\columnwidth]{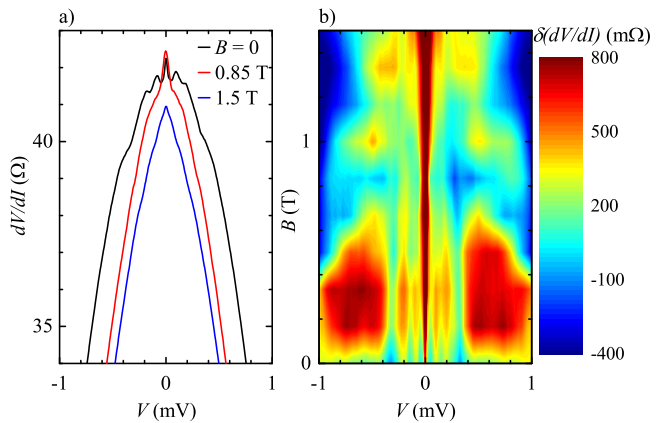}
\caption{(Color online) (a) Examples of $dV/dI(V)$ characteristics for different magnetic fields at low (30~mK) temperature. The resonances disappear completely above  0.85~T, while the zero-bias $dV/dI$ peak survives. At higher fields, $dV/dI(V)$ curve still demonstrates  non-trivial shape, with non-zero first derivative at zero bias (b) Colormap showing evolution of $\delta (dV/dI(V))$ with magnetic field. The oscillations' positions are constant up to their disappearance above  0.85~T. Additional features appears in higher fields.}
\label{IV_magn}
\end{figure}

Thus, the observed resonances requires relation to surface states at the Nb-$WTe_2$ interface.  WTe$_2$ is regarded as type-II Weyl semimetal~\cite{wang,mazhar,wang-miao} hosting topological Fermi arcs  states on (001) surfaces~\cite{soluyanov}, which was demonstrated in several experiments~\cite{wu,wang}. In this case, an analog of Tomasch oscillations is allowed for transport along the topological surface state across the region of proximity-induced superconductivity near the niobium superconducting lead~\cite{adroguer,melnikov}. 

The resonances appear as Fabry-Perot-type transmission resonances for Bogoliubov quasiparticles in a long $l>>\xi$ single NS junction, $\xi$ is a coherence length.  They are situated~\cite{adroguer,melnikov}  at energies $eV_n=\sqrt{\Delta_{Ind}^2+\left(hv_fn/2l\right)^2}$, where $n=0,1,2...$, $v_F$ is the Fermi velocity. According to this relation, the induced gap $\Delta_{ind}$ is reflected by the  zero-bias structure~\cite{heslinga,klapwijk17,ingasb} in Fig.~\ref{IVb}. By fitting the resonance positions, we obtain   $\Delta_{ind}\approx 0.1$~meV,  the effective junction dimension as $l\approx 2\mu$m for $v_F\sim 1.5\cdot10^5 \frac{m}{s}$ (from ARPES data\cite{bruno}).

\begin{figure}
\includegraphics[width=0.7\columnwidth]{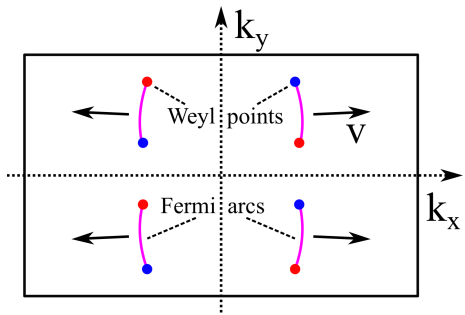}
\caption{(Color online) Sketch of Fermi arcs and projections of Weyl points on (001) surface Brillouin zone. WTe$_2$ has eight Weyl points which are connected by for Fermi arcs. In general, eight Weyl points are connected by four Fermi arcs on each surface, these arcs  define two directions for carriers in the surface state.  In WTe$_2$, the Weyl point are aligned along the $b$-axis~\cite{li} of the crystal, so these two directions merge into one along the $a$-axis, which defines  a specific transmission direction of carriers.
 }
\label{Arcs}
\end{figure}

 This value is comparable with the  dimensions of the planar Nb-$WTe_2$ junction (below $10 \mu$m), taking in mind  that placing $WTe_2$ on top of Nb does not guarantee good contact at all the surface. $l\sim 2\mu$m is also about the mean free path along the a-axis in $WTe_2$. It much exceeds the coherence length~\cite{kulik-long,dubos} $l>>\xi=(l_e \times \hbar v_F^N/\pi\Delta_{in})^{1/2}\approx 200$~nm, which is obligatory to observe Tomasch oscillations  for transport along the topological surface state~\cite{adroguer,melnikov}. 

 The crucial point is that transmission resonances implies well-defined junction length $l$. For the planar NS junction without axial symmetry, (nearly rectangular in our case), $l$ is different in different directions, which should smear the resonances for trivial surface states,  e.g., originating from band bending near the WTe$_2$ surface.  On the other hand, Weyl surface states  inherit the chiral property of the Chern insulator edge states~\cite{armitage}, where the preferable directions is defined by  Fermi arcs on a particular crystal surface.  In WTe$_2$, the Weyl point are aligned along the $b$-axis~\cite{li} of the crystal,   forming preferable directions for surface carriers along the $a$-axis, see Fig.~\ref{Arcs}.  Thus, observation of sharp subgap resonances is specific for topological transport within the Fermi arc surface states at the Nb-$WTe_2$ interface. Since the resonances are defined by the interference effects, they are obviously suppressed in magnetic field above 0.85~T see Fig.~\ref{IV_magn}. However,  the $dV/dI(V)$ curve evolution in higher fields is unusual for Andreev reflection and requires further investigations.

\section{Conclusion}

As a conclusion, we experimentally investigate charge transport through the interface between a niobium superconductor and a three-dimensional WTe$_2$ Weyl semimetal. In addition to classical Andreev reflection, we observe sharp non-periodic subgap resistance resonances. From an analysis of their positions, magnetic field and temperature dependencies, we can interpret them as an analog of Tomasch oscillations for transport along the topological surface state across the region of proximity-induced superconductivity at the Nb-WTe$_2$ interface.  Observation of distinct geometrical resonances implies a specific transmission direction for carriers, which is a hallmark of the Fermi arc surface states.

\acknowledgments

We wish to thank V.T.~Dolgopolov, V.A.~Volkov, I.~Gornyi, and A.S.~Melnikov for fruitful discussions, and S.S~Khasanov for X-ray sample characterization.  We gratefully acknowledge financial support by the RFBR (projects 16-02-00405 and 18-02-00368) and RAS.

\end{document}